\documentclass{moriond}

\def\be{\begin{equation}}
\def\ee{\end{equation}}
\def\bea{\begin{eqnarray}}
\def\eea{\end{eqnarray}}

\usepackage{amsmath}
\usepackage{comment}
\usepackage{wrapfig}

\newcommand{\Photo}{}

\begin{document}
\vspace*{4cm}
\title{Low Energy Calibration in DUNE’s Far Detector prototypes}

\author{Emile Lavaut -On behalf of the DUNE collaboration}

\address{High Energy Physics Department, Laboratoire Irene Juliot-Curie, 15 rue Georges Clémenceau Orsay, France}

\maketitle\abstracts{
The Deep Underground Neutrino Experiment (DUNE) is a next-generation long-baseline neutrino experiment. In addition to GeV-scale oscillation measurements ($\delta_{CP}$, $\theta_{23}$ octant, mass ordering), DUNE features a low-energy (MeV-scale) program targeting solar, supernova burst (SNB), and Diffuse Supernova Background (DSNB) neutrinos. Accurate reconstruction and background understanding are critical. $^{39}$Ar $\beta$-decays, naturally present in LAr, provide a uniform background and can be used for calibration. This proceeding presents an analysis of isolated MeV-scale energy deposits in the ProtoDUNE-HD (PDHD) prototype. Using cosmic + beam data (run 28086), we analyze energy spectra and spatial distributions of $^{39}$Ar, $^{232}$Th, and $^{207}$Bi. A calibration factor $c_A = (3.9 \pm 0.3)\times 10^{-2}~\text{MeV}/\text{ADC} \times \text{tick}$ and recombination factor $R = 0.60 \pm 0.05$ are extracted, consistent with expectations. The $^{207}$Bi source is spatially resolved with cm-level precision, and $^{232}$Th hot spots align with the field cage structure, offering further calibration potential. These results demonstrate PDHD’s capability for MeV-scale reconstruction, supporting DUNE's low-energy physics goals.}

\section{Introduction}

\subsection{DUNE detector}

The Deep Underground Neutrino Experiment (DUNE) supports a robust low-energy (MeV-scale) neutrino program alongside its GeV-scale oscillation goals \cite{DUNE2020_Physics}. It aims to detect supernova neutrinos with high statistics, offering insights into black hole formation and contributing to multi-messenger astronomy with a 3.3° angular resolution for a 10 kpc SN \cite{DUNE2024}. DUNE also seeks to observe the elusive Diffuse Supernova Neutrino Background (DSNB) and solar \textit{hep} neutrinos to probe both the standard solar model (SSM) and non-standard neutrino interactions. The Far Detector complex (four 17-ktLArTPCs, 1.5 km underground), offering a low cosmic ray rate, is the perfect place to perform this low-energy physics\cite{DUNE2020}. However, success depends on precise calibration, reconstruction algorithms, and stable detector operation, including LAr purity and noise control.

\subsection{DUNE prototypes}
\begin{figure}[ht]
\begin{minipage}{0.48\linewidth}
\centerline{\includegraphics[width=0.7\linewidth]{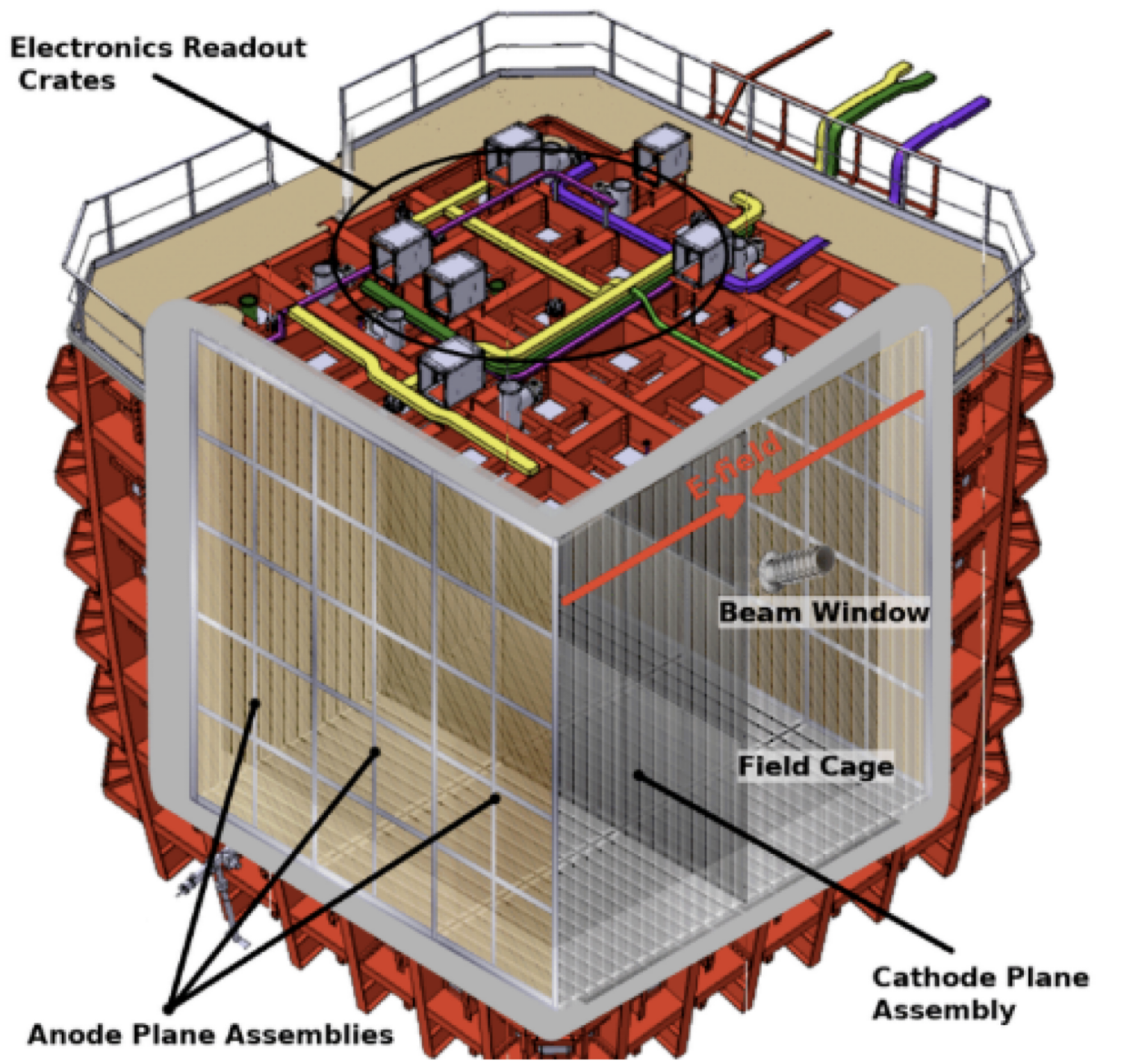}}
\end{minipage}
\hfill
\begin{minipage}{0.48\linewidth}
\centerline{\includegraphics[width=0.9\linewidth]{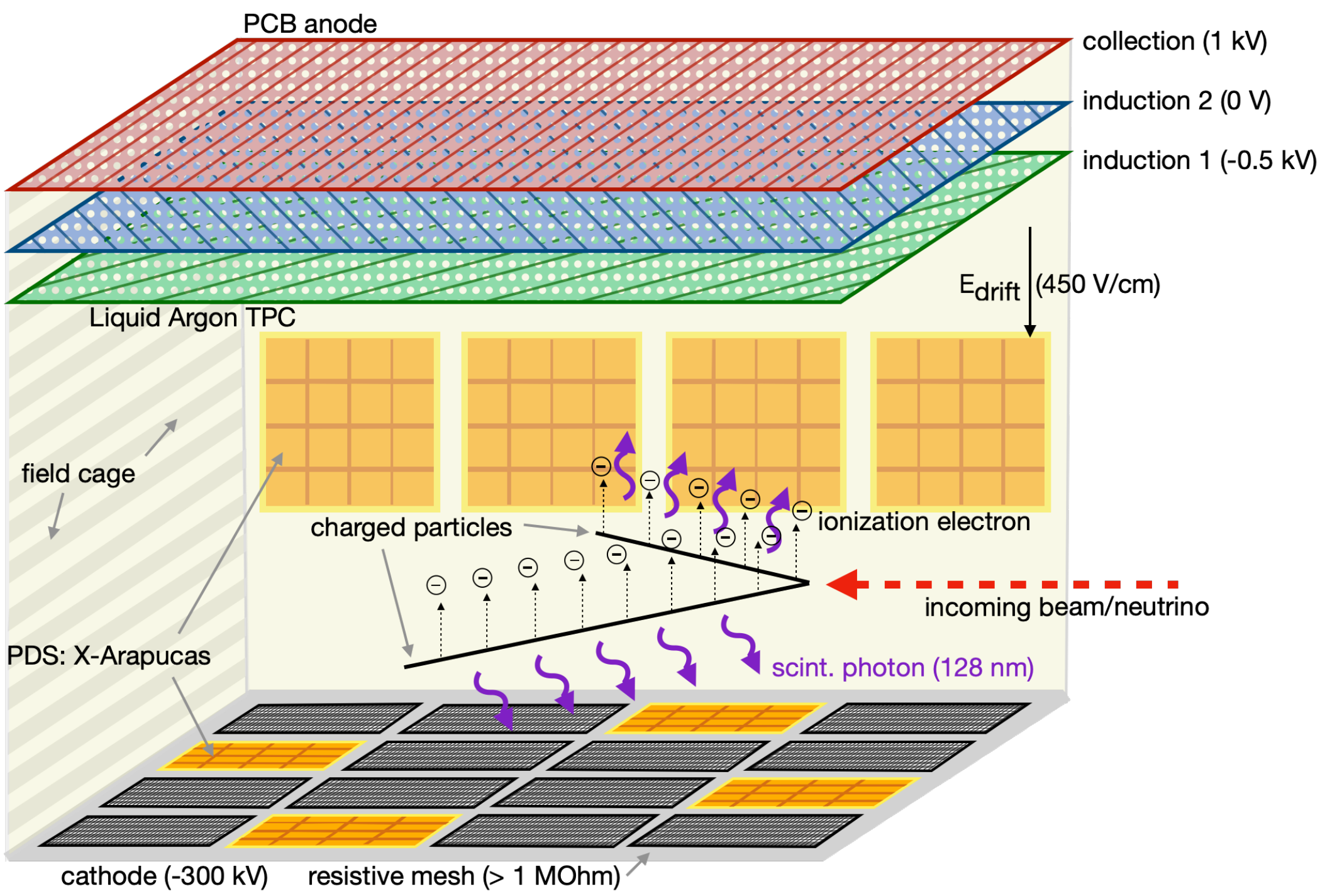}}
\end{minipage}
\caption[]{PDHD detector diagram showing its different components from \cite{DUNE2020} (left) and diagram of a neutrino/beam event in LArTPC (right)}
\label{fig:PDHD_TPC}
\end{figure}

ProtoDUNE Horizontal Drift (PDHD) and ProtoDUNE Vertical Drift (PDVD) are two prototypes of the DUNE Far Detector, each with an active volume of about 200 tons, located at the CERN Neutrino Platform. As shown in Fig.~\ref{fig:PDHD_TPC}, they are designed to validate DUNE FD technologies, including charge reconstruction and low-energy calibration. Due to their surface location, cosmic muon backgrounds are significant, making MeV-scale studies more challenging. PDHD collected data from May to November 2024, while PDVD is scheduled to begin data-taking in May 2025. 

\subsection{Low-Energy Calibration Sources}

In the few MeV range, natural radioactivity is a key background in the DUNE Far Detector (FD), with dominant contributions from $^{39}\text{Ar}$ $\beta$-decays—roughly $10^7$ per second in the 40 kton volume. Despite a low Q-value (565 keV), these decays are abundant and can aid in low-energy calibration, liquid argon purity checks, and understanding recombination or space-charge effects thanks to their uniform distribution. Other internal sources like $^{85}\text{Kr}$, $^{40}\text{K}$, $^{222}\text{Rn}$, and $^{232}\text{Th}$ contribute less frequently but at higher energies. Their localized origin helps in spatial reconstruction studies and additional calibration. Neutrons and gammas from cavern rock also contribute to the background.

In the surface-based ProtoDUNE-HD (PDHD), similar background sources are observed but at lower rates—around $10^5$ $^{39}\text{Ar}$ decays per second, because of the reduced volume of the prototype. This study focuses on reconstructing and identifying $^{39}\text{Ar}$ events. Also, the 2.6 MeV gamma line from the $^{232}\text{Th}$ chain, linked to the field cage and cathode structure, is studied. Additionally, an artificial $^{207}\text{Bi}$ source has been installed for calibration purposes. These sources give point-like signature signals in an environment full of cosmic tracks, which makes them challenging to reconstruct: here is the first analysis made with the dedicated \texttt{SingleHit} tool in \texttt{LArSoft} software.

\section{Results}
\begin{figure}[t]
\begin{minipage}{0.33\linewidth}
\centerline{\includegraphics[width=0.8\linewidth]{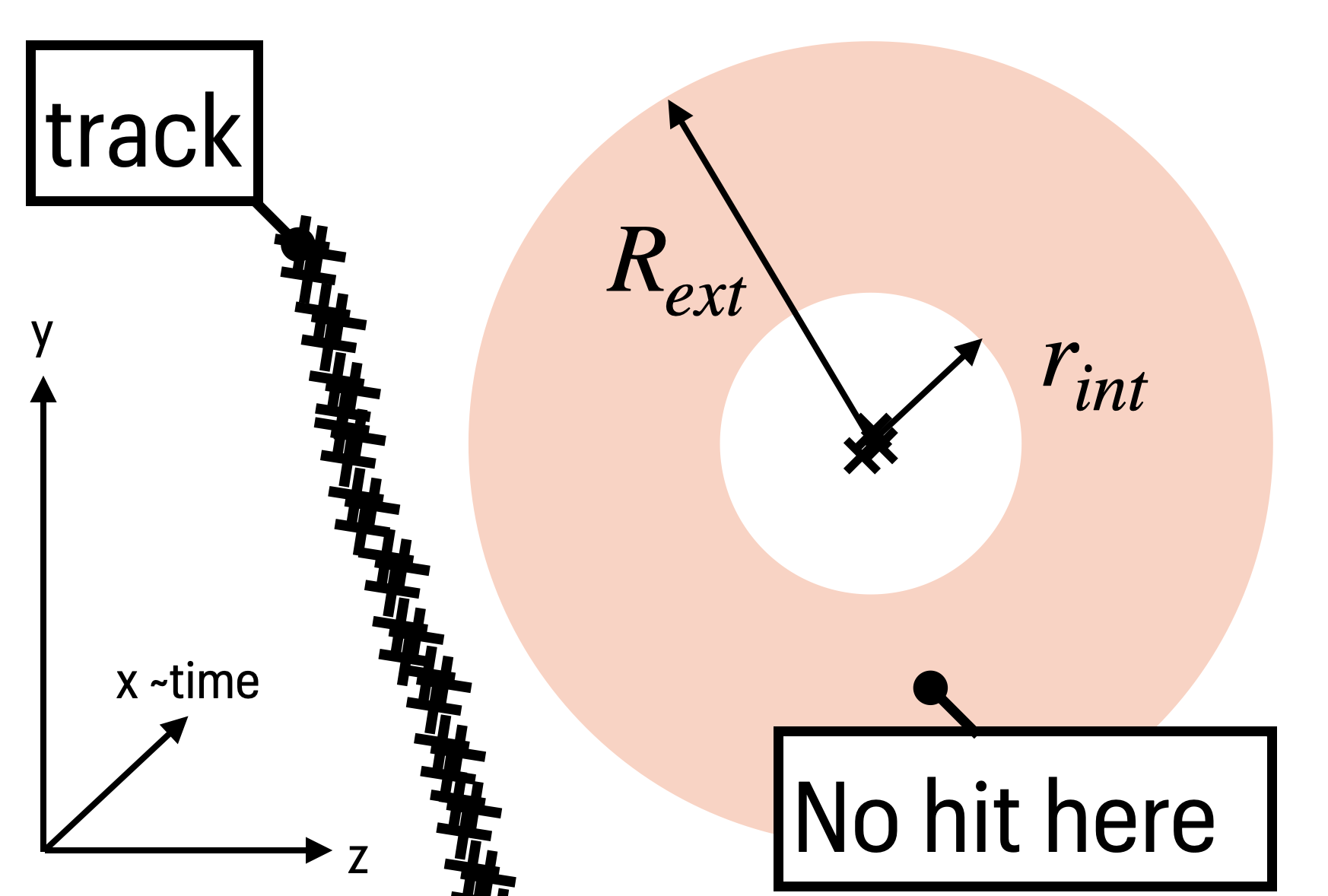}}
\end{minipage}
\vline
\begin{minipage}{0.33\linewidth}
\centerline{\includegraphics[width=0.9\linewidth]{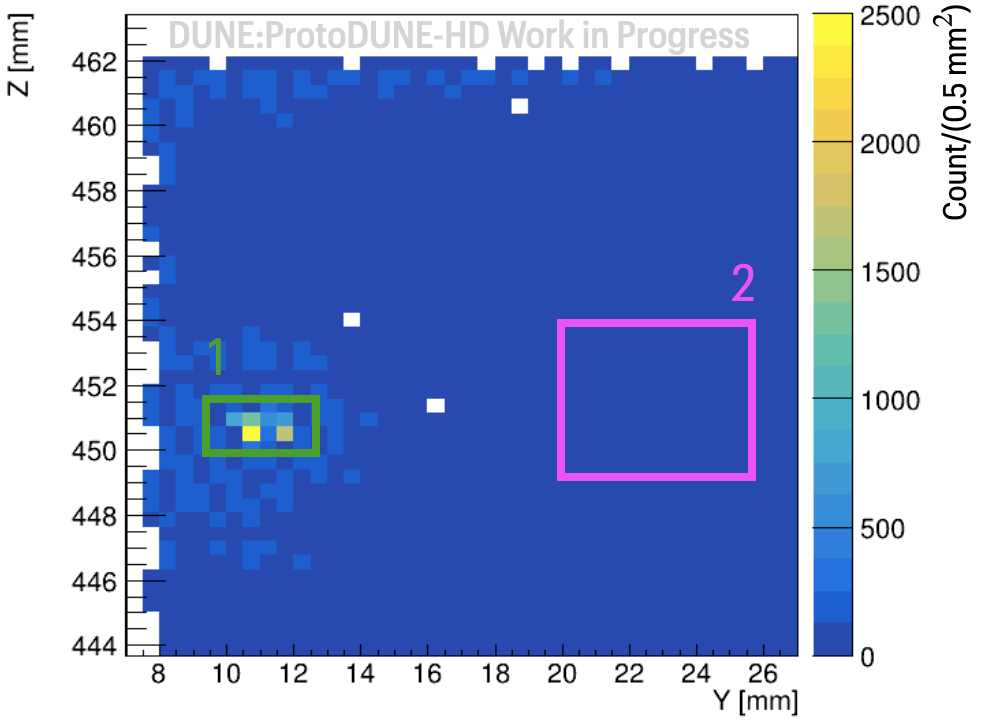}}
\end{minipage}
\begin{minipage}{0.33\linewidth}
\centerline{\includegraphics[width=0.8\linewidth]{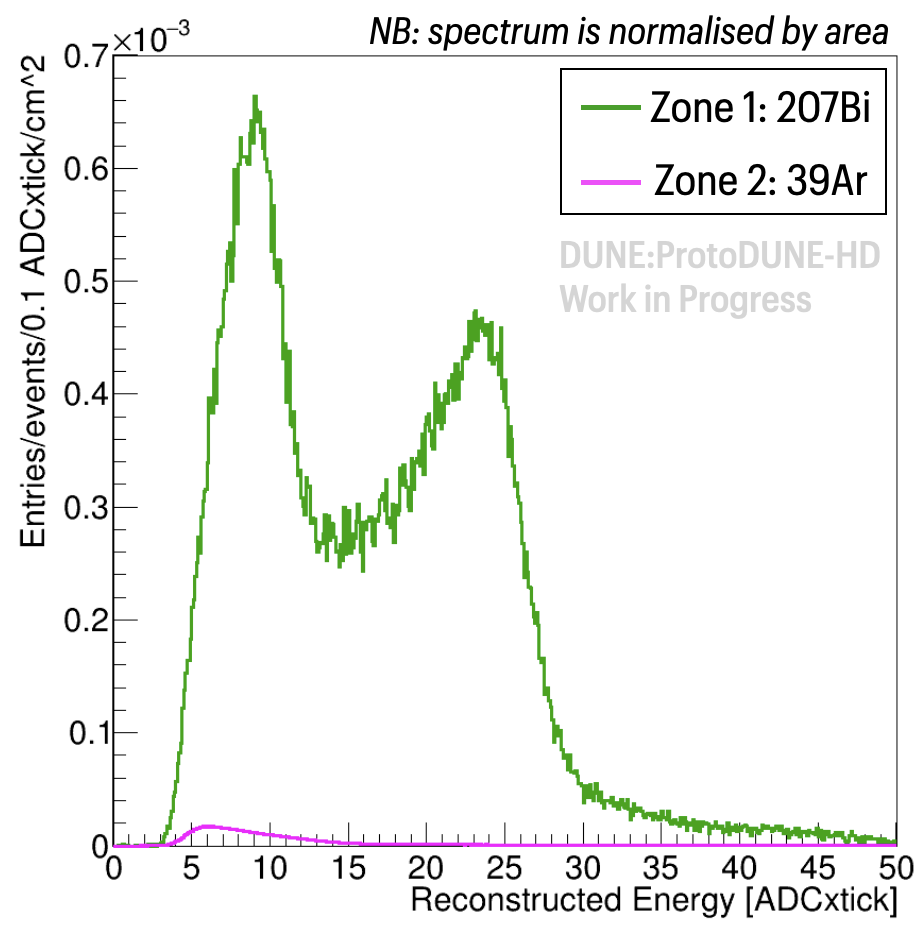}}
\end{minipage}
\caption[]{Diagram showing how \texttt{SingleHit} selects radioactive-like signals (left), the black crosses symbolize hits. Right: Spatial identification of the Bismuth source (area 1) and associated spectrum in green, we identify conversion electron peaks and the Compton front of the bismuth source. In area 2 (pink) is the normalized $^{39}\text{Ar}$ signal.}
\label{fig:SingleHit_bi}
\end{figure}

All the results presented here focus on PDHD ionization signal (charge), either for the official DUNE Monte Carlo simulation or run number 28086 for the data. This run operated with an electric field of $500~\text{V/cm}$, at 87 K, with a purity greater than 30~ms; it is a 36~min long run with cosmic muons and 1~GeV electron beam taken on July 21st, 2024.
The data and Monte Carlo are analyzed using a custom analysis framework: the \texttt{SingleHit} tool. This tool's goal is to identify, select, and cluster radioactive decays ($^{39}\text{Ar}$, $^{207}\text{Bi}$, $^{232}\text{Th}$), and suppress cosmic muons and beam tracks. In the end, the signal is a little blip (few hits with a spatial expansion of a few cm) isolated in the volume (far from tracks); see Fig.~\ref{fig:SingleHit_bi} -left-.

\begin{figure}[b]
\begin{minipage}{0.33\linewidth}
\centerline{\includegraphics[width=0.8\linewidth]{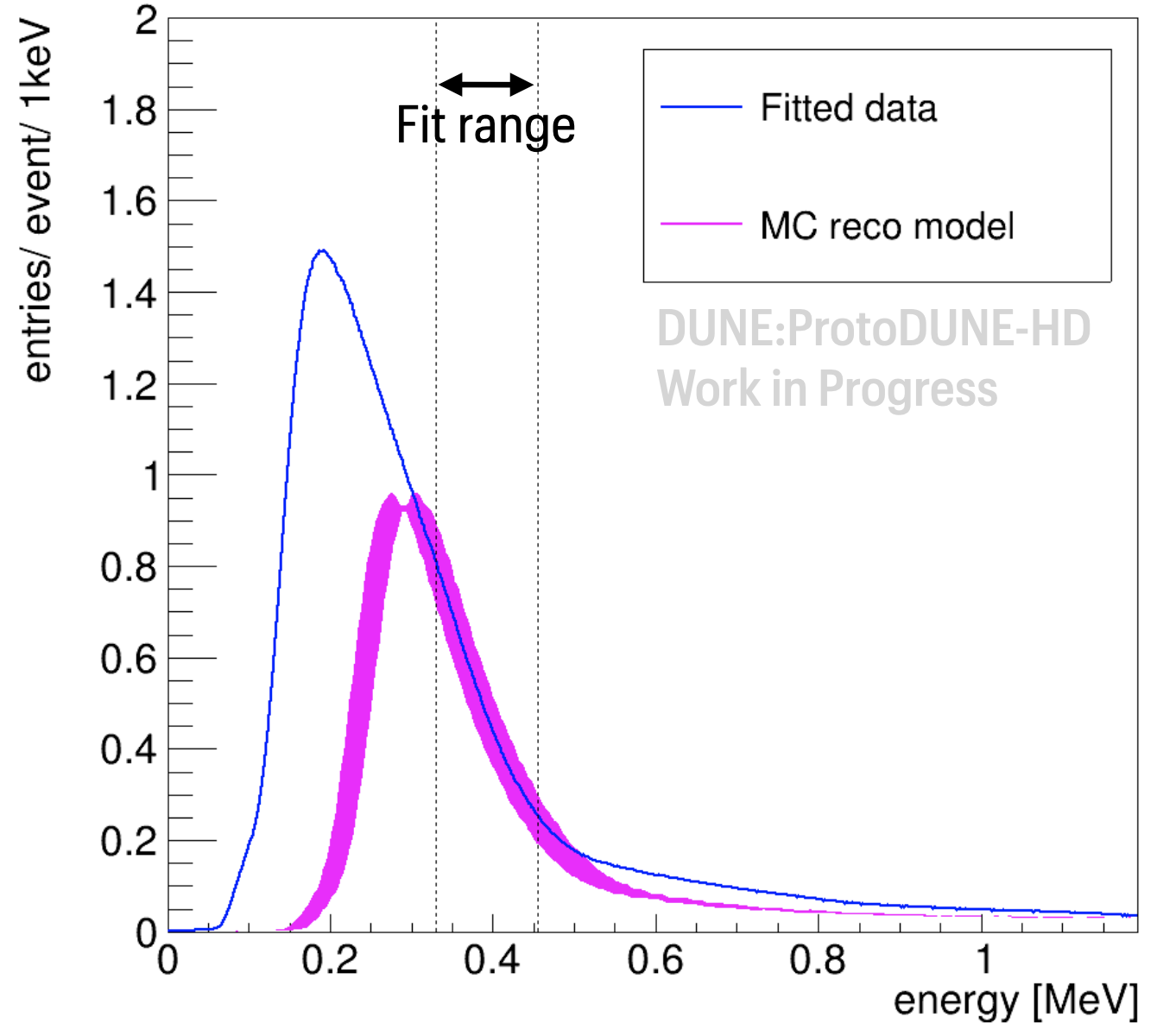}}
\end{minipage}
\vline
\begin{minipage}{0.33\linewidth}
\centerline{\includegraphics[width=0.7\linewidth]{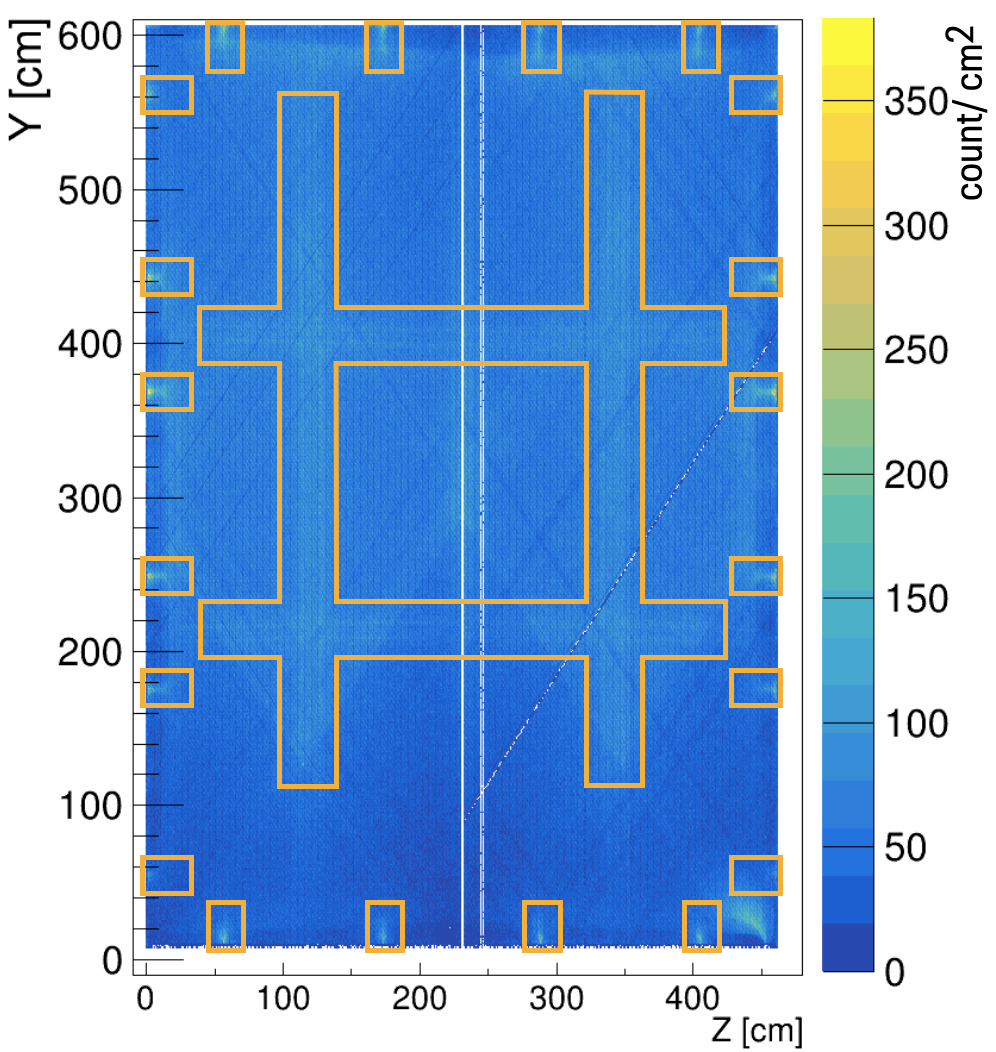}}
\end{minipage}
\begin{minipage}{0.33\linewidth}
\centerline{\includegraphics[width=1.1\linewidth]{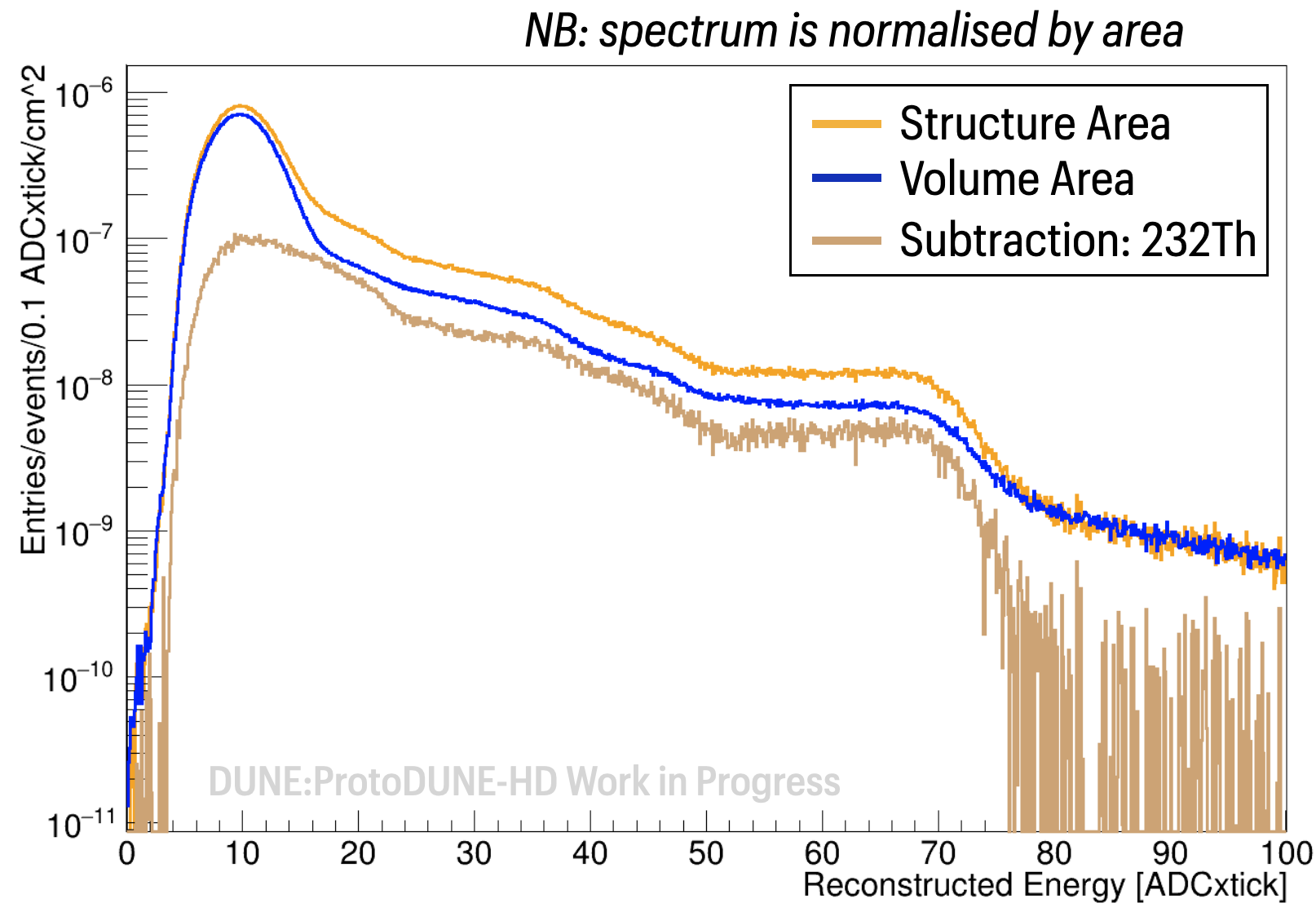}}
\end{minipage}
\caption[]{End of $^{39}\text{Ar}$ induced peak is fitted with reconstructed MC simulation (left). The MC model embodies some systematic errors due to the use of the modified Box Model in the simulation, which is known for being inexact at this energy scale. Right: spatial identification of the cathode and field-cage structure and definition (orange line) of the structure area. The spectrum of structure area (orange) subtracted by volume area (area outside of the orange line, in blue) shows a peak at $\sim$70 ADC$\times$tick identified as the 2.6 MeV Th gamma line's Compton front.}
\label{fig:ar_th}
\end{figure}

\begin{wrapfigure}{R}{0.6\textwidth}
  \begin{center}
    \includegraphics[width=0.6\textwidth]{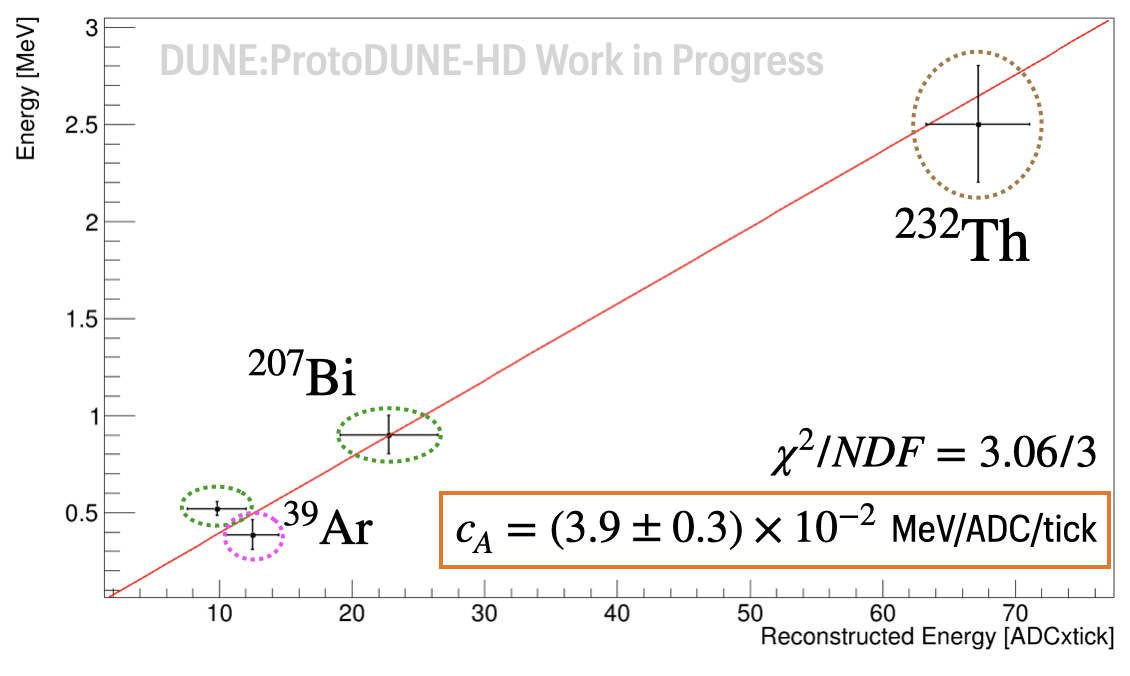}
  \end{center}
  \caption{Calibration curve obtained with PDHD data after identification of radiological sources}\label{fig:calib}
\end{wrapfigure}

An analysis of the official DUNE Monte-Carlo simulation (cosmic muons, 1 GeV electron beam, $^{39}\text{Ar}$, $^{85}\text{Kr}$ and $^{222}\text{Rn}$) has been performed with \texttt{SingleHit}. Systematic errors need to be incorporated into the MC model because the modified Box Model \cite{ICARUS2004} has been used as the recombination model and is known for being inexact in this energy range. The reconstructed Monte-Carlo is then used to perform a calibration fit on the reconstructed data, Fig.~\ref{fig:ar_th} -left. A calibration factor for $^{39}\text{Ar}$ only, $c_A = 0.031 \pm 0.001~\text{MeV/ADC/tick}$ is found, it is in the same order of magnitude as the estimated one at higher energy $c_A^{\mathrm{HE}} \approx 0.036 \text{ MeV/ADC/tick}$. On Fig.~\ref{fig:calib} the corresponding calibration point is tagged $^{39}\text{Ar}$.

Then the $^{207}\text{Bi}$ source can be studied, the position of the source is reconstructed precisely at the centimeter level, which is the detector resolution, as shown in Fig.~\ref{fig:SingleHit_bi} -middle-. In this spatial distribution, the Bismuth area is identified (green area 1), then the energy distribution of the cluster in this area is plotted in Fig.~\ref{fig:SingleHit_bi} -right- (in green). To compare, a test area is defined (pink area 2), and the energy distribution of this area is also plotted in Fig.~\ref{fig:SingleHit_bi} -right- (in pink). From this plot, we can identify the 570 keV and the 976 keV electron conversion peaks, with the associated Compton fronts due to gamma scattering. On Fig.~\ref{fig:calib} the corresponding calibration points are tagged $^{207}\text{Bi}$.

The last signal studied here is the 2.6~MeV gamma line from the Thallium at the end of the $^{232}\text{Th}$ chain. When the spatial distribution of reconstructed low-energy clusters is plotted for the full detector, see Fig.~\ref{fig:ar_th} -middle-, various features appear: a "cross" structure in the middle, and hot-spots on the side, top and bottom edges of the detector (these features are delimited by the orange line). These features are at the exact position (at the centimeter level) of the cathode structure for the "cross" and the field-cage beams for the hot spots. As for the $^{207}\text{Bi}$, a fiducialization is applied (orange line) and the energy distribution is plotted, orange histogram on Fig.~\ref{fig:ar_th} -right-. Then, the distribution from the outside of this fiducialization is subtracted (blue histogram on Fig.~\ref{fig:ar_th} -right-) to obtain the brown histogram. The peak around $\sim$70~ADC$\times$tick is identified as the Compton front from the 2.6~MeV gamma line of the Thallium because the material of the structure (field-cage beam and cathode structure) is known to bring some Thorium contribution in the detector \cite{MicroBooNE}. On Fig.~\ref{fig:calib} the corresponding calibration point is tagged $^{232}\text{Th}$.

\subsection{Calibration Factor and recombination}

In order to analyze and interpret the low-energy spectrum in ProtoDUNE-HD, it is essential to estimate the calibration factor, $f$, which connects the collected charge to the deposited energy through the relation:  
$E~[\text{MeV}] = f \times Q~[\text{ADC} \times \text{tick}]$, where $E$ is the true deposited energy and $Q$ is the collected ionization charge. From Fig.~\ref{fig:calib} a calibration factor is fitted (assuming linear calibration): $c_A = (3.9 \pm 0.3) \times 10^{-2}~\text{MeV}/(\text{ADC} \times \text{tick})$. This calibration factor is consistent with one estimated at higher energies: $c_A^{\mathrm{HE}} \approx 0.036 \text{ MeV/ADC/tick}$.

The parameter $R$, modeling the recombination, can be extracted from this calibration factor. The recombination is an important feature of LArTPCs: energy deposition ionizes the liquid argon, resulting in a cloud of ions $Ar^+$ and a cloud of electrons. The cloud of electrons is the signal that drifts towards the anode. But some of the electrons will recombine instantaneously with the surrounding ions: it is the recombination. $R$ is the ratio of the visible energy to the true deposited energy, it is well modelled for track-like events (for example, the Modified Box Model) but not at the MeV scale. The relation:  $R= \frac{W_{\text{ions}}}{c_A \times g_e}$, with $W_{\text{ions}}=23.6 \times 10^{-6}~\text{MeV}/e^-$ the ionization energy and $g_e=10^{-3}~\text{ADC} \times \text{ticks}/e^-$ the electronics gain factor, gives $R=0.60 \pm 0.05$. This value is consistent with other experiments \cite{ICARUS2004} and has the same order of magnitude for the error.

\section{ Conclusion and Outlook}

This study demonstrates the capability of ProtoDUNE-HD to reconstruct and calibrate MeV-scale energy deposits using natural and artificial radioactive sources. The newly developed \texttt{SingleHit} tool effectively isolates low-energy signals from cosmic and beam backgrounds, enabling precise spatial and energy analysis. Calibration using $^{39}$Ar, $^{207}$Bi, and $^{232}$Th yielded a consistent calibration factor and recombination parameter, validating the low-energy response. The cm-level source localization and spectral features confirm detector performance and spatial reconstruction accuracy. These results support DUNE's low-energy program and readiness for supernova and solar neutrino detection.

\end{document}